\documentclass[aps,prc,reprint,superscriptaddress,longbibliography]{revtex4-2}

\usepackage{graphicx}   % Include figure filed
\usepackage{float}
\usepackage{xspace}     % Include xspace
\usepackage{url}
\usepackage[mathlines]{lineno}% Enable numbering of text and display math
\usepackage{bm}
\usepackage{amsmath}
\usepackage{xcolor}
\usepackage[driverfallback=dvipdfm]{hyperref}
\usepackage{ulem}
\usepackage{mathtools}
\usepackage{amsmath}

\newcommand{\pt}{\mbox{$p_\mathrm{T}$}\xspace}
\newcommand{\myeta}{\mbox{$\eta$}\xspace}

\newcommand{\sqsnn}{\mbox{$\sqrt{s_{\mathrm{NN}}}$}\xspace}
\newcommand{\rn}{\mbox{$r_{\mathrm{n}}$}\xspace}

\newcommand{\myetaRef}{\mbox{$\eta_\text{ref}$}\space}

\usepackage{tikz,xcolor,hyperref}

\definecolor{lime}{HTML}{A6CE39}
\DeclareRobustCommand{\orcidicon}{
	\begin{tikzpicture}
	\draw[lime, fill=lime] (0,0) 
	circle [radius=0.16] 
	node[white] {{\fontfamily{qag}\selectfont \tiny ID}};
	\draw[white, fill=white] (-0.0625,0.095) 
	circle [radius=0.007];
	\end{tikzpicture}
	\hspace{-2mm}
}
\foreach \x in {A, ..., Z}{%
	\expandafter\xdef\csname orcid\x\endcsname{\noexpand\href{https://orcid.org/\csname orcidauthor\x\endcsname}{\noexpand\orcidicon}}
}
\foreach \x in {A, ..., Z}{%
	\expandafter\xdef\csname orcid\x\endcsname{\noexpand\href{https://orcid.org/\csname orcidauthor\x\endcsname}{\noexpand\orcidicon}}
}

\begin{document}

\title{Collision energy and system size dependence of longitudinal flow decorrelation in heavy-ion collisions at RHIC energies}

\author{Gaoguo Yan\orcidA{}}\affiliation{School of Physics and Electronic Engineering, Linyi University, Linyi, 276000, China}\affiliation{Institute of Frontier and Interdisciplinary Science, Shandong University, Qingdao, 266237, China}
\author{Maowu Nie\orcidB{}{}}\email[]{maowu.nie@sdu.edu.cn}\affiliation{Institute of Frontier and Interdisciplinary Science, Shandong University, Qingdao, 266237, China}\affiliation{Key Laboratory of Particle Physics and Particle Irradiation, Ministry of Education, Shandong University, Qingdao, Shandong, 266237, China}
\author{Zhenyu Chen\orcidC{}}\email[]{zhenyuchen@sdu.edu.cn}\affiliation{Institute of Frontier and Interdisciplinary Science, Shandong University, Qingdao, 266237, China}\affiliation{Key Laboratory of Particle Physics and Particle Irradiation, Ministry of Education, Shandong University, Qingdao, Shandong, 266237, China} 
\author{Li Yi\orcidD{}}\email[]{li.yi@sdu.edu.cn}\affiliation{Institute of Frontier and Interdisciplinary Science, Shandong University, Qingdao, 266237, China}\affiliation{Key Laboratory of Particle Physics and Particle Irradiation, Ministry of Education, Shandong University, Qingdao, Shandong, 266237, China} 
\author{Jiangyong Jia\orcidE{}}\email[]{jiangyong.jia@stonybrook.edu}\affiliation{Department of Chemistry, Stony Brook University, Stony Brook, New York 11794, USA}\affiliation{Physics Department, Brookhaven National Laboratory, Upton, New York 11976, USA} 

\begin{abstract}
In heavy-ion collisions, the initial collision geometry and its fluctuations drive the collective expansion of final-state hadrons in the transverse plane. However, longitudinal fluctuations induce event-plane twist and flow magnitude asymmetries, collectively known as longitudinal flow decorrelation. Using a multi-phase transport (AMPT) model, we systematically investigate the dependence of collision energy and system size of this phenomenon with Au+Au collisions at \sqsnn = 19.6, 27, 54.4, 200 GeV and isobar collisions (Zr+Zr and Ru+Ru) at \sqsnn = 200 GeV. The results reveal two distinct decorrelation components: $r_n(\eta)$, which includes flow magnitude asymmetry and event-plane twist, and $R_n(\eta)$ which arises purely from event-plane twist. Both $r_n(\eta)$ and $R_n(\eta)$ decrease linearly with $\eta$ and exhibit a significant dependence on collision energy and the size of the system. Through the slope parameters $F_n$ in the linear parametrization $r_n(\eta) = 1-2F_n\eta$, we can quantify the strength of decorrelation. We further observe that both $F_2$ and $F_3$ demonstrate a pronounced power-law scaling behavior with collision energy, following the relation $F_n \propto log \sqrt{s_{NN}}$. These results provide valuable insights into the three-dimensional modeling of the initial stage and the evolution of relativistic heavy-ion collisions.
\end{abstract}

\maketitle

%%%%%%%%%%%%%%%%%%%%%%%%%%%%%%%%%%%%%%%%%%%%%%%%%%%%%%%%%%%%%%%%%%%%%%%%%%%%%%%%%%%
%%%%%%%%%%%%%%%%%%%%%%%%%%%%%%%%%%%%%%%%%%%%%%%%%%%%%%%%%%%%%%%%%%%%%%%%%%%%%%%%%%%
\section{Introduction}
Anisotropic flow~\cite{Ollitrault:1992bk}, which describes the azimuthal angle distribution of final-state charged hadrons, provides critical constraints on the initial state and transport properties of the Quark-Gluon Plasma (QGP) created in high-energy heavy-ion collisions at the Relativistic Heavy Ion Collider (RHIC) and the Large Hadron Collider (LHC). The flow can be characterized through Fourier expansion of hadron yield distribution in azimuthal angle $\phi$~\cite{Ollitrault:1993ba,Voloshin:1994mz,Poskanzer:1998yz}:
\begin{equation}
    dN/d\phi \propto 1+2\sum\nolimits_{n}v_{n}\,\mathrm{cos}\, n(\phi-\psi_{n})
\end{equation} 
\label{Eq1: vn}
\noindent 
where $v_{n}$ and $\psi_{n}$ represent the magnitude and phase of the $\mathrm{n^{th}}$-order of flow vector, respectively. Here, $v_2$ denotes the elliptic flow and $v_3$ is the triangular flow. The absence of sine terms in this expansion arises from symmetry constraints with respect to the event plane. Extensive measurements of the magnitudes of $v_n$ and event-by-event fluctuations have been performed at RHIC~\cite{STAR:2000ekf,STAR:2004jwm,STAR:2013qio,PHENIX:2004jsa,PHOBOS:2006dbo} and LHC~\cite{ALICE:2010suc,ALICE:2011ab,CMS:2013wjq,ATLAS:2012at}. With a boost invariant based space-time evolution of heavy ion collisions scenario, the (2+1)D event-by-event viscous hydrodynamical model has achieved great success in understanding these anisotropic flow parameters~\cite{Gale:2013da}. It is now well known that the higher order harmonics can better constrain viscosity and fluctuating initial conditions, and the temperature dependence of $\eta/s$ can be well handled by the rapidity differential anisotropic flow in heavy-ion collisions~\cite{Denicol:2015nhu}. These measurements are essential for extracting QGP properties through comparisons with hydrodynamic and transport models. However, recent theoretical~\cite{Bozek:2010vz, Xiao:2012uw, Huo:2013qma, Bozek:2015bha, Jia:2014ysa} and experimental progresses~\cite{PHOBOS:2004nvy} suggest that the boost invariant approximation scenario is not accurate enough, as the two-particle correlations as a function of pseudorapidity revealed strong event-by-event fluctuations, i.e. $V_{n\Delta}(\eta_1, \eta_2) \neq v_n(\eta_1)v_n(\eta_2)$. 

Many previous studies have shown that this non-boost invariant can lead to longitudinal flow decorrelation. For example, torqued fireball model~\cite{Bozek:2015bna}, 3DMCG model~\cite{Shen:2017bsr}, 3DGlasma~\cite{Schenke:2016ksl} and the AMPT model~\cite{Pang:2015zrq,Wu:2018cpc}. The different geometry of wounded nucleons between the forward and backward rapidity directions leads to a twist in the final state event-plane angles or an asymmetry in the flow magnitudes due to random fluctuation of participating nucleons. The signature of longitudinal flow decorrelation has been measured by experiments at RHIC~\cite{Nie:2019bgd,Yan:2023ugh} and LHC~\cite{CMS:2015xmx,ATLAS:2017rij,ATLAS:2020sgl}.

Flow decorrelation can be quantified with a flow decorrelation observable, factorization ratio $r_n$, 
\begin{align} 
r_n(\eta, \eta_{ref}) = \frac{V_{n\Delta}(-\eta,\eta_{ref})}{V_{n\Delta}(\eta,\eta_{ref})}
\end{align}
\label{Eq2: rn}

where $\eta_{ref}$ is the reference pseudorapidity common to the numerator and the denominator, thus $r_n$ is sensitive to the correlation between $\eta$ and $-\eta$. If the value of \rn is lower than unity which means the presence of longitudinal flow decorrelation due to the factorization breaks down between $-\myeta$ and $\myeta$.

Based on Ref.\cite{Jia:2017kdq}, the observable $r_{n}(\eta)$ can be approximately described by a linear function of \myeta with a negative slope: 
\begin{equation}
        r_{n}(\eta) \approx 1-2F_{n}\eta, F_{n}=F_{n}^{\mathrm{asy}}+F_{n}^{\mathrm{twi}}
        \label{EQfn}
\end{equation}
\label{Eq3: Fn}
The longitudinal decorrelation mainly includes contribution from asymmetry in the magnitude of $v_n$, $F_{n}^{\mathrm{asy}}$ , and the twist of the event plane, $F_{n}^{\mathrm{twi}}$, between \myeta and -\myeta. To estimate the separate contribution of the asymmetry and twist effect, a new observable is used:
\onecolumngrid
\begin{equation}
    \begin{aligned}
        R_{n}(\eta) &= \frac{
            \langle q_{n}(-\eta_{\text{ref}})q^{*}_{n}(\eta)q_{n}(-\eta)q^{*}_{n}(\eta_{\text{ref}}) \rangle
        }
        {
            \langle q_{n}(-\eta_{\text{ref}})q^{*}_{n}(-\eta)q_{n}(\eta)q^{*}_{n}(\eta_{\text{ref}}) \rangle
        } \\
        &= \frac{
            \langle v_{n}(-\eta_{\text{ref}})v_{n}(-\eta)v_{n}(\eta)v_{n}(\eta_{\text{ref}})
            \cos\left\{ n[\psi_{n}(-\eta_{\text{ref}})-\psi_{n}(\eta_{\text{ref}})+(\psi_{n}(-\eta)-\psi_{n}(\eta))] \right\} \rangle
        }
        {
            \langle v_{n}(-\eta_{\text{ref}})v_{n}(-\eta)v_{n}(\eta)v_{n}(\eta_{\text{ref}})
            \cos\left\{ n[\psi_{n}(-\eta_{\text{ref}})-\psi_{n}(\eta_{\text{ref}})-(\psi_{n}(-\eta)-\psi_{n}(\eta))] \right\} \rangle
        }
    \end{aligned}
\end{equation}
\label{Eq4: Rn}
\twocolumngrid

The choice of the range of $\eta_{\text{ref}}$ is fully motivated by physical considerations. A smaller gap between $\eta$ and $\eta_{\text{ref}}$ can lead to sizable nonflow contributions, mainly from dijets. The reference pseudorapidity dependence $\eta_{\text{ref}}$ has already studied experimentally~\cite{Yan:2023ugh}, $3.1 < |\myetaRef| < 5.1$ reduces the effect of dijets and provides good statistical precision at RHIC energies. A recent study also suggests the $\eta_{\text{ref}}$ dependence has been attributed to local fluctuations in rapidity~\cite{Jia:2024xvl}. In a previous AMPT study~\cite{Dixit:2023bho}, the energy dependence of $r_2$ and $r_3$ at RHIC energies was investigated using a pseudorapidity range $2.1 < |\myetaRef| < 5.1$. However, the results—especially for $r_2$—may have been significantly affected by nonflow contributions. To achieve a more robust energy dependence, it is crucial to effectively subtract the contributions of nonflow effects. Moreover, the system size dependence of flow decorrelation is less explored at RHIC energies. Additionally, a comprehensive study on $R_n$ is still needed to probe initial state geometry fluctuations.

In this paper, we present a systematic study on longitudinal flow decorrelation using the AMPT (A Multi-Phase Transport) model, analyzing Au+Au collisions at \sqsnn = 19.6, 27, 54.4 and 200 GeV, along with isobar (Zr+Zr and Ru+Ru) collisions at \sqsnn = 200 GeV. The collision energy dependence can be systematically investigated through comparisons of Au+Au collisions across different energies, while the system size dependence is quantified by comparing Au+Au and isobar collisions. This multi-dimensional approach enables simultaneous characterization of both energy- and system size-dependent of flow decorrelation.
%%%%%%%%%%%%%%%%%%%%%%%%%%%%%%%%%%%%%%%%%%%%%%%%%%%%%%%%%%%%%%%%%%%%%%%%%%%%%%%%%%%
%%%%%%%%%%%%%%%%%%%%%%%%%%%%%%%%%%%%%%%%%%%%%%%%%%%%%%%%%%%%%%%%%%%%%%%%%%%%%%%%%%%
\section{Model setup}
The AMPT model with string melting scenario~\cite{Lin:2004en} is utilized to simulate Au+Au collision at \sqsnn = 19.6, 27, 54.4 and 200 GeV, isobar (Zr+Zr and Ru+Ru) collisions at \sqsnn = 200 GeV. In string melting version of AMPT, Monte Carlo Glauber model \cite{Miller:2007ri} is used to provide the initial conditions, HIJING model \cite{Wang:1991hta} generates the initial partons by strings and mini-jet melting that will follow elastic parton cascade, simulating by ZPC model \cite{Zhang:1997ej}, eventually, hadronization and hadron rescatterings are described by the quark coalescence model and ART model \cite{Li:1995pra}, respectively. The effect of longitudinal flow decorrelation arise from generating varying string lengths of initial partons of the interaction resulting in fluctuations in the initial geometry along the longitudinal direction \cite{Pang:2015zrq, Wu:2018cpc, Ma:2016fve}. The elastic parton-parton cross section is chosen as the standard value of 3 mb at the top RHIC energy to maintain parameter consistency across different collision energies.

The observable, $r_{n}(\myeta)$, for flow decorrelations in Eq.(2) is constructed using final state hadrons with transverse momentum $0.4 < \pt < 4$ GeV and pseudorapidity $|\myeta| < 1.5$ in centrality bins, where centralities are determined by the multiplicity distribution of charged particles within $|\myeta| < 0.5$ that is chosen for corresponding to the acceptance of the Time Projection Chamber (TPC) detector in the STAR experiment. In experimental measurements, a rapidity gap is often required between $\eta$ and $\eta_{\mathrm {ref}}$ to suppress non-flow correlations associated with jet fragmentation and resonance decays. For this analysis, we choose the reference pseudorapidity to be $3.1 < |\eta_{\mathrm {ref}}| < 5.1$ for $r_2$ and $2.1 < |\eta_{\mathrm {ref}}| < 5.1$ for $r_3$. As the isobar collisions have certain nuclear deformation, the Wood-Saxon parameters for Zr+Zr and Ru+Ru are set with $R_0$ = 5.09 and a = 0.52 fm, the deformation parameter is set with $\beta_2$ = 0.162 and $\beta_3$ = 0 for Ru, $\beta_2$ = 0.06 and $\beta_3$ = 0.2 for Zr, same as the previous study~\cite{Nie:2022gbg}.
%%%%%%%%%%%%%%%%%%%%%%%%%%%%%%%%%%%%%%%%%%%%%%%%%%%%%%%%%%%%%%%%%%%%%%%%%%%%%%%%%%%
%%%%%%%%%%%%%%%%%%%%%%%%%%%%%%%%%%%%%%%%%%%%%%%%%%%%%%%%%%%%%%%%%%%%%%%%%%%%%%%%%%%
\section{Results and discussions}
\begin{figure}[H]
    \centering
    \includegraphics[width=0.98\linewidth]{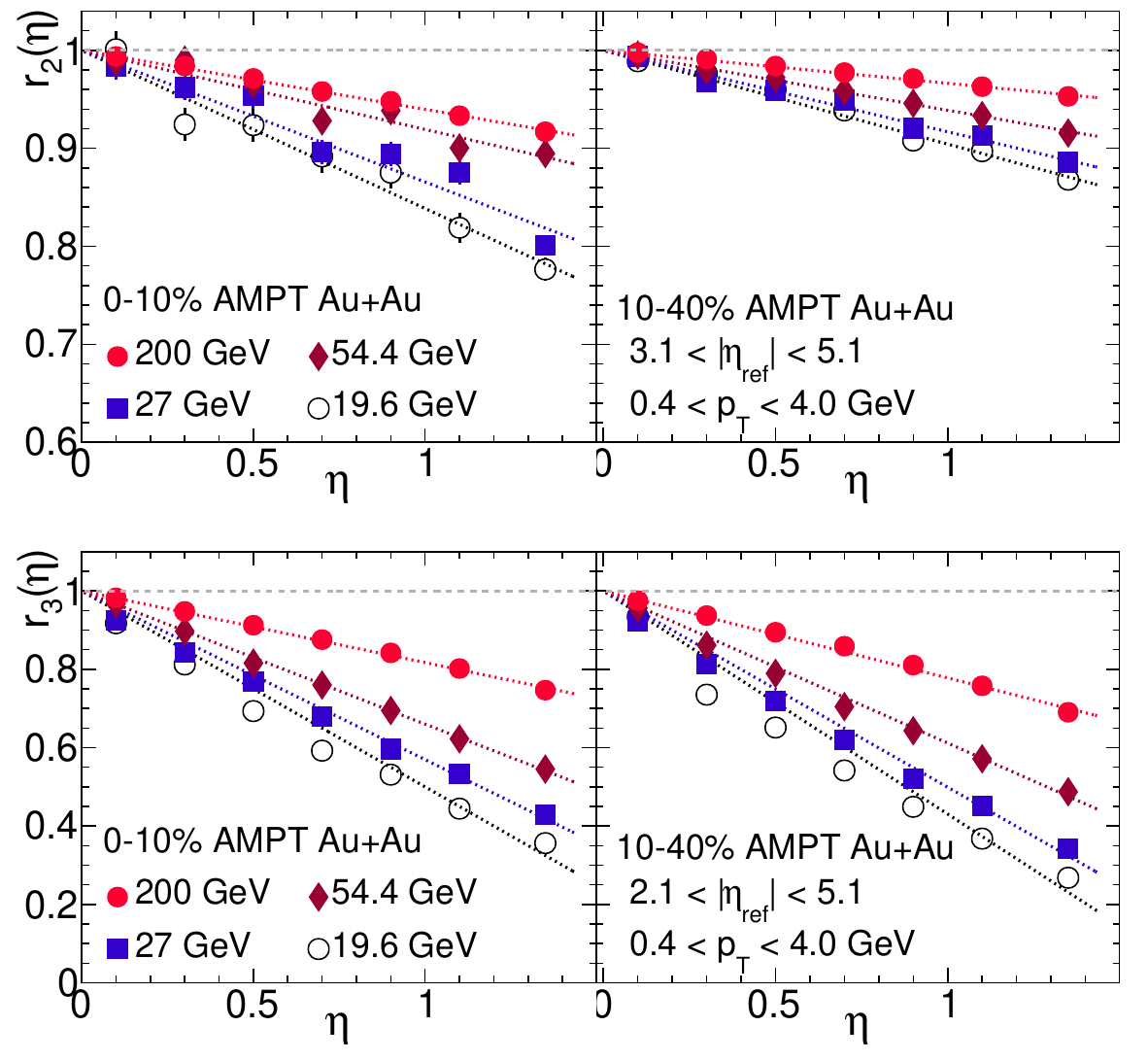}
    \caption{The $r_n(\eta)$ (n=2, 3) compared between the Au+Au collisions at \sqsnn = 19.6 (black), 27 (red), 54.4 (blue) and 200 (orange) GeV in centrality bins: 0-10\% and 10-40\% by simulating AMPT model. The dashed line represents a linear fit to the data.}
    \label{fig_rn_energy}
\end{figure}

Figure \ref{fig_rn_energy} depicts $r_{2}$ and $r_{3}$ as a function of \myeta for Au+Au collisions in 0-10\% and 10-40\% centrality intervals at four collision energies. Both $r_{2}$ and $r_{3}$ decrease linearly with increasing \myeta. The decreasing trend can be described by a linear fit (dash line). The values of $r_{2}$ and $r_{3}$ both decrease from 200 GeV to 19.6 GeV, which indicates that lower energy leads to larger flow decorrelation due to a less boost invariant.

\begin{figure}[htbp]
    \centering
    \includegraphics[width=0.98\linewidth]{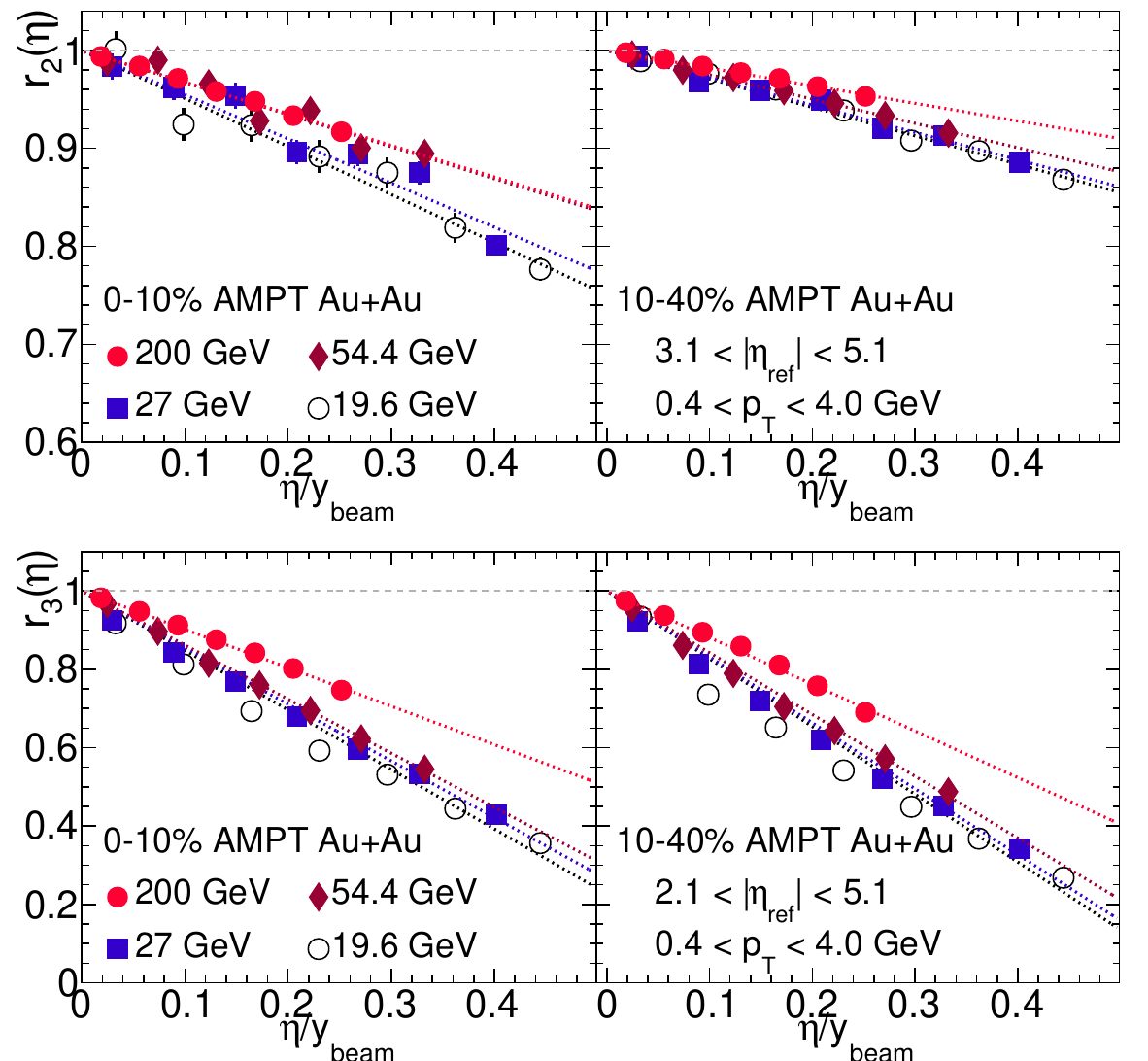}
    \caption{The $r_n(\eta/y_{beam})$ (n=2, 3) compared between the Au+Au collisions at \sqsnn = 19.6 (black), 27 (red), 54.4 (blue) and 200 (orange) GeV in centrality bins: 0-10\% and 10-40\% by simulating AMPT model. The dashed line represents a linear fit to the data.}
    \label{fig_rn_scale}
\end{figure}

To account for the beam-rapidity dependence, a rapidity normalization procedure is further applied for the comparison. Figure \ref{fig_rn_scale} shows $r_{2}$ and $r_{3}$ as a function of normalized pseudorapidity $\eta/y_{beam}$, where $y_{beam}$ = 5.36 for 200 GeV, $y_{beam}$ = 4.06 for 54.4 GeV, $y_{beam}$ = 3.36 for 27 GeV and $y_{beam}$ = 3.04 for 19.6 GeV. After normalizing to the beam rapidity, both second- and third-order flow decorrelations exhibit a clear dependence on collision energy, which indicates the nontrivial dynamical behavior cannot be fully explained by simple scaling with the beam rapidity. At 200 GeV, the flow decorrelation is weaker compared to other collision energies. For the remaining energies, the results are generally consistent, except for the second-order decorrelation in the 0-10\% centrality range. Notably, the decorrelation strengths at 27 GeV and 19.6 GeV tend to align within uncertainties, suggesting a possible hint of non-linear dependence on collision energy.

%=======================================
% Figure1: energy dependence
%=======================================
%=======================================
%=======================================
% Figure2: energy dependence by scale beam rapidity
%=======================================
%=======================================

\begin{figure}[H]
    \centering
    \includegraphics[width=0.98\linewidth]{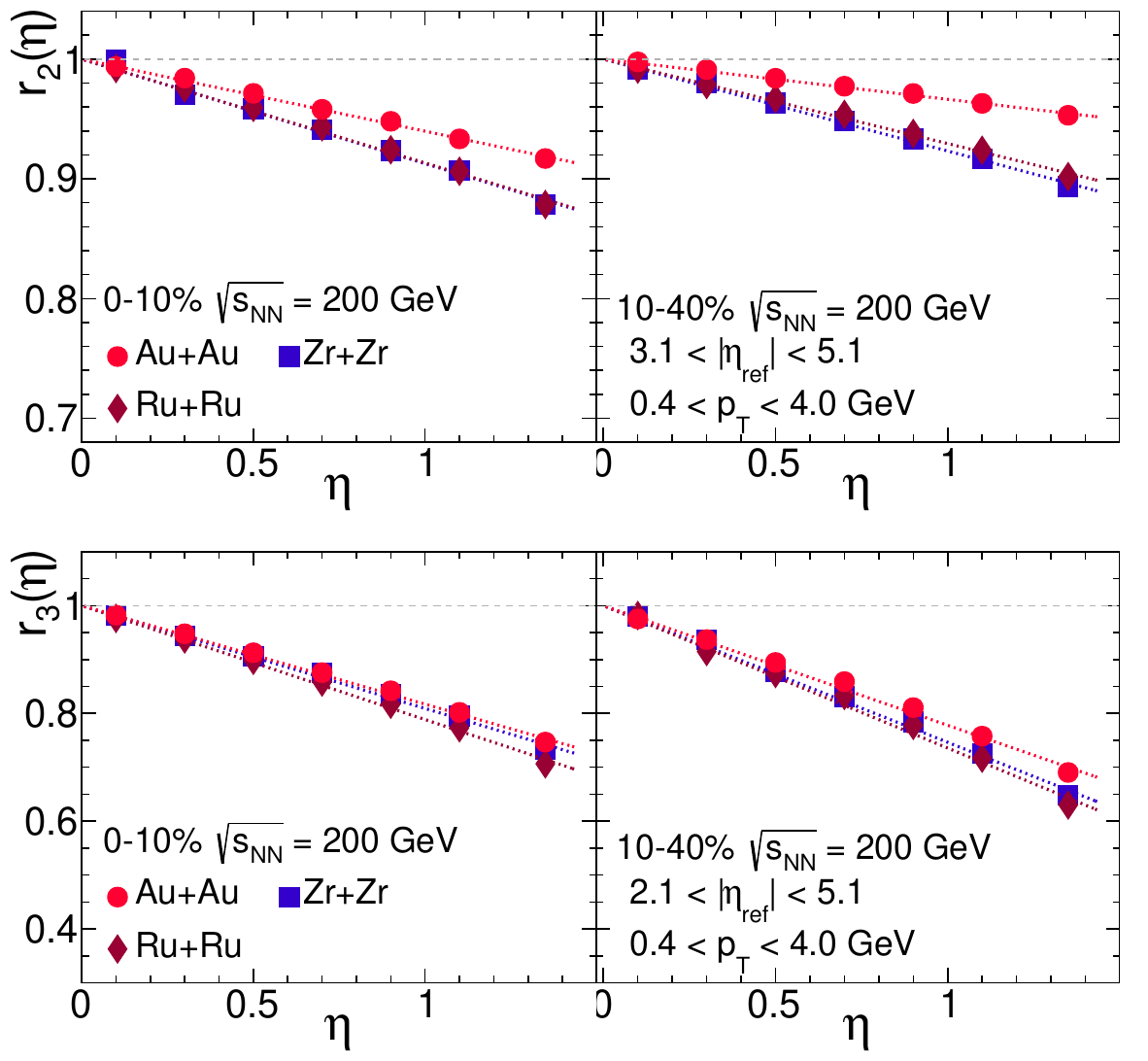}
    \caption{The $r_n(\eta)$ (n=2, 3) compared between the Zr+Zr (red), Ru+Ru (blue), Au+Au (Orange) collisions at \sqsnn = 200 GeV in centrality bins: 0-10\% and 10-40\% by simulating AMPT model. The vertical line on the data points represent the statistical uncertainty.}
    \label{fig_rn_system}
\end{figure}

Figure \ref{fig_rn_system} compares the decorrelation observables $r_{2}$ and $r_{3}$ in Au+Au and isobar collisions (Zr+Zr and Ru+Ru) at $\sqrt{s_{\mathrm{NN}}} = 200$ GeV. Au has a larger system size (in terms of nucleon number) than Zr and Ru. 
The overall level of $v_2$  decorrelation in Au+Au collisions is considerably larger than in isobar collisions, where $r_{2}$ values in Au+Au are approximately 4\% larger than in isobar collisions in 0-10\% centrality and 5\% larger in 10-40\% centrality. Meanwhile, the difference in $r_{3}$ between Au+Au and isobar collisions is relatively small, with a difference of around 1\% in 0-10\% centrality and a 2\% difference in the 10-40\% centrality. 
The results indicate a stronger decorrelation in smaller collision systems (Zr+Zr and Ru+Ru) compared to larger systems (Au+Au). This is consistent
%=======================================
% Figure3: system size dependence
%=======================================
%=======================================
\noindent with the picture where smaller initial system size is associated with more fluctuations due to fewer participant nucleons, leading to a larger difference between forward and backward pseudorapidity directions. The tiny difference between Zr+Zr and Ru+Ru is attributed to differences in nuclear structure~\cite{Nie:2022gbg}.
%=======================================
%To study the centrality dependence of flow decorrelation, \rtwo and \rthree are parameterized with a linear function as shown in the Eq.(\ref{EQfn}). Therefore, the slope parameters, $F_2$ and $F_3$, directly quantify the strength of the decorrelation effect. Figure \ref{fig_fnEnergy} shows $F_2$ and $F_3$ as a function of $N_{\mathrm{part}}$ for the four collision energies. The values of $F_2$ first decrease and then increase as a function of $N_{\mathrm{part}}$. It is consistent with the fact that $v_2$ is dominant by the initial geometry, where $v_2$ is largest in mid-central collisions with less geometry fluctuations, and become small in central and peripheral collisions with larger geometry fluctuations. All the four energies show similar centrality dependence. A clear hierarchy is observed, where 19.6 GeV has the largest $F_2$ for almost all centralities. While $F_3$ shows weak centrality dependence, the reason is consistent with the picture $v_3$ is fluctuation driven, and less dependent on centralities. The $F_3$ values also show clear energy dependence. The energy dependence of flow decorrelation is consistent with the scenario where lower collision energies result in a smaller number of initial partons and shorter string lengths, leading to stronger decorrelation~\cite{Pang:2015zrq}.

\begin{figure}[htbp]
	\centering
	\includegraphics[width=\linewidth]{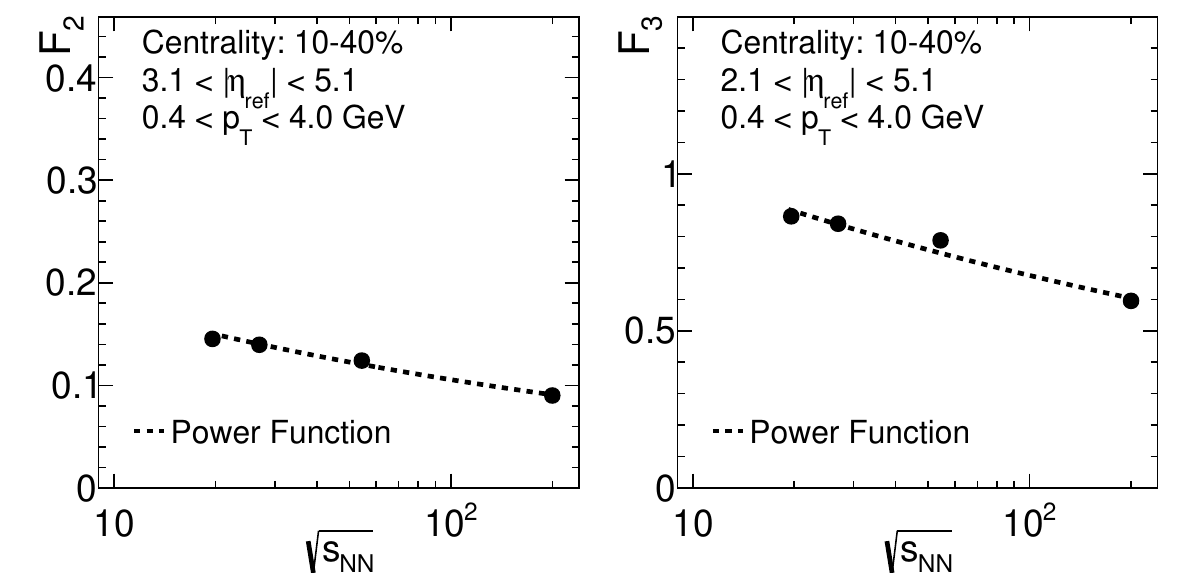}
	\caption{The slope parameter $F_2$ and $F_3$ are plotted as a function of collision energies.}
	\label{fig_fnEnergy}
\end{figure}

To directly quantify the strength of flow decorrelation,  the slope parameters $F_2$ and $F_3$ were extracted from the parameterization described in Eq.~(\ref{EQfn}) within 10--40\% centrality interval. These parameters are plotted as functions of collision energy, as shown in Fig.~\ref{fig_fnEnergy}. A clear hierarchy is observed, where 19.6 GeV has the largest $F_2$ and $F_3$. The energy dependence of flow decorrelation is consistent with the scenario where lower collision energies result in a smaller number of initial partons and shorter string lengths, leading to stronger decorrelation~\cite{Pang:2015zrq}. We also find that both $F_2$ and $F_3$ exhibit a clear power-law dependence on collision energy,  where $F_n$ $\propto$ $log \sqrt{s_{NN}}$. This observation should be further validated with experimental data.

%=======================================
% Figure4: Fn
%=======================================
%\begin{figure}[htbp]
%    \centering
%    \includegraphics[width=\linewidth]{Fn_energy.pdf}
%    \caption{The slope parameter $F_2$ and $F_3$ are plotted as function of participant at \sqsnn = 19.6 (black), 27 (red), 54.4 (blue) and 200 (orange) GeV in Au+Au collisions. The vertical line on the data points represent the statistical uncertainty.}
%    \label{fig_fnEnergy}
%\end{figure}

\begin{figure}[htbp]
    \centering
    \includegraphics[width=\linewidth]{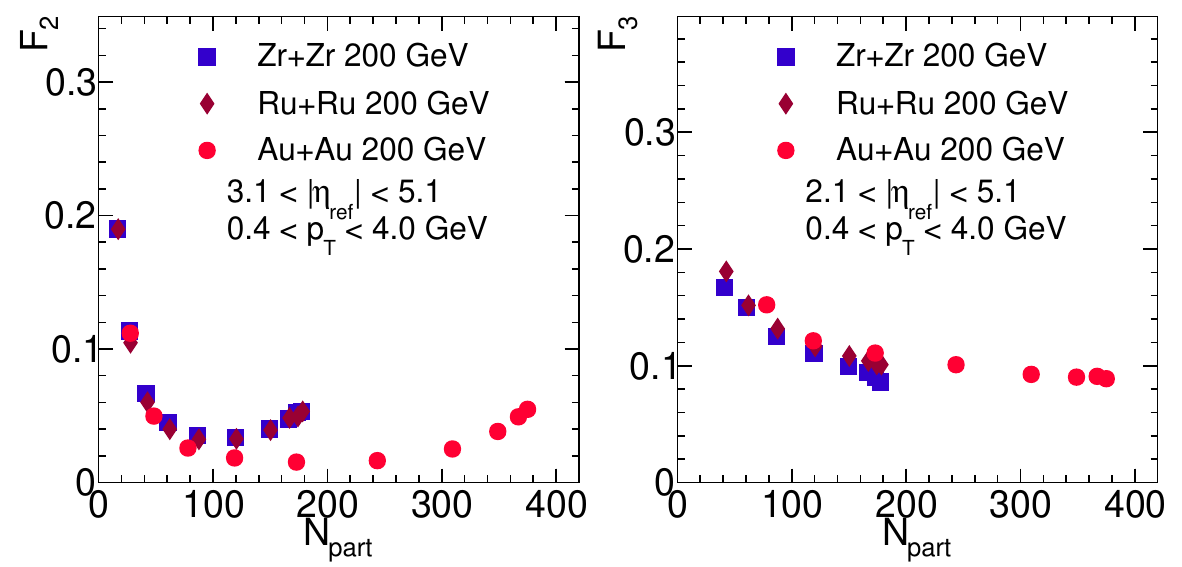}
    \caption{The slope parameter $F_2$ and $F_3$ are plotted as function of participant at \sqsnn = 200 GeV in Au+Au (orange), Zr+Zr (red) and Ru+Ru (blue) collisions. The vertical line on the data points represent the statistical uncertainty.}
    \label{fig_fnSystem}
\end{figure}

Figure~\ref{fig_fnSystem} shows $F_2$ and $F_3$ as a function of $N_{\mathrm{part}}$ for Au+Au and isobar collisions. $F_2$ and $F_3$ for Ru+Ru and Zr+Zr almost overlap with each other. The value of $F_2$ in isobar collisions larger than in Au+Au collisions at same $N_{\mathrm{part}}$. The results suggest clear system size dependence. While for $F_3$, Au+Au results are slightly larger than isobar collisions at same $N_{\mathrm{part}}$. The opposite trend of system size dependence between $F_2$ and $F_3$ is already observed by ATLAS experiments~\cite{ATLAS:2020sgl}, and the results are also consistent with the Glauber model study. The $N_{\mathrm{part}}$ and $N_{\mathrm{part}}/2A$, where A is the atomic number, can be considered as measured the absolute and scaled system size. 

\begin{figure}[htbp]
    \centering
    \includegraphics[width=\linewidth]{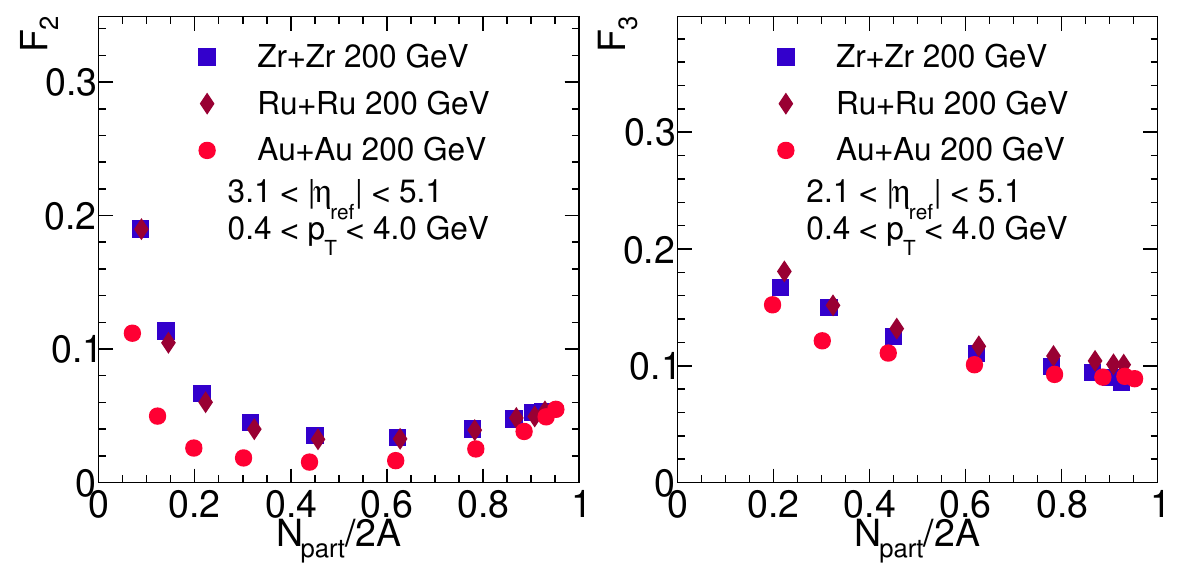}
    \caption{The slope parameter $F_2$ and $F_3$ are plotted as function of scaled participant at \sqsnn = 200 GeV in Au+Au (orange), Zr+Zr (red) and Ru+Ru (blue) collisions. The vertical line on the data points represent the statistical uncertainty.}
    \label{fig_fnSystem2A}
\end{figure}

Previous studies indicate that $\varepsilon_2$ for different systems primarily governed by the deterministic geometry of the overlapping region, as evidenced by the universal scaling observed when plotted against $N_{\mathrm{part}}$/2A. In contrast, $\varepsilon_3$ is driven by the random fluctuations of quark constituents. This is supported by the observed universal scaling at small $N_{\mathrm{part}}$ (where fluctuations dominate), which transitions to a system-dependent deviation at large $N_{\mathrm{part}}$~\cite{Behera:2020mol}. We have further plotted $F_2$ and $F_3$ as a function of $N_{\mathrm{part}}/2A$ for Au+Au and isobar collisions as shown in Figure~\ref{fig_fnSystem2A}. The opposite trend of system size dependence observed for $F_2$ and $F_3$ appears to vanish in contrast to longitudinal eccentricity decorrelations. Thus the system size dependence serves as a powerful constraint, enabling a clearer disentanglement of the longitudinal structure of the initial state in heavy-ion collisions, and can be verified by RHIC-STAR experimental data.
%=======================================
% Figure5: Fn
%=======================================
%=======================================
% Figure6: Fn
%=======================================
%=======================================

\begin{figure}[htbp]
    \centering
    \includegraphics[width=\linewidth]{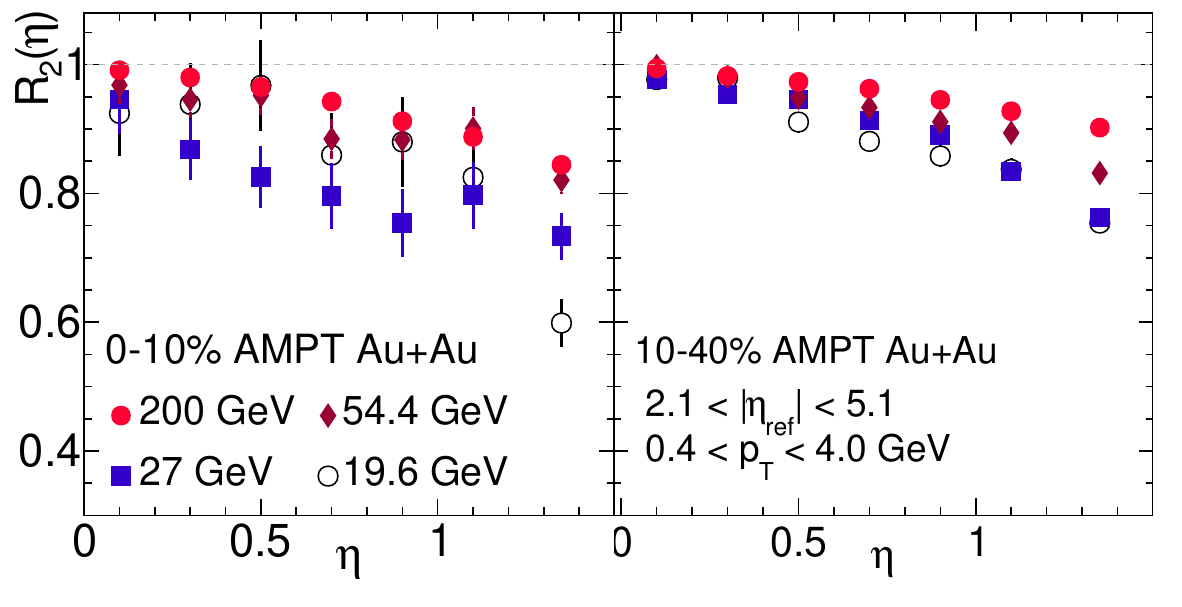}
    \caption{The $R_2(\eta)$ compared at \sqsnn = 19.6 (black), 27 (red), 54.4 (blue) and 200 (orange) GeV in Au+Au collisions in centrality 0-10\% and 10-40\%. The vertical line on the data points represent the statistical uncertainty}
    \label{fig_R2Energy}
\end{figure}

\begin{figure}[htbp]
    \centering
    \includegraphics[width=\linewidth]{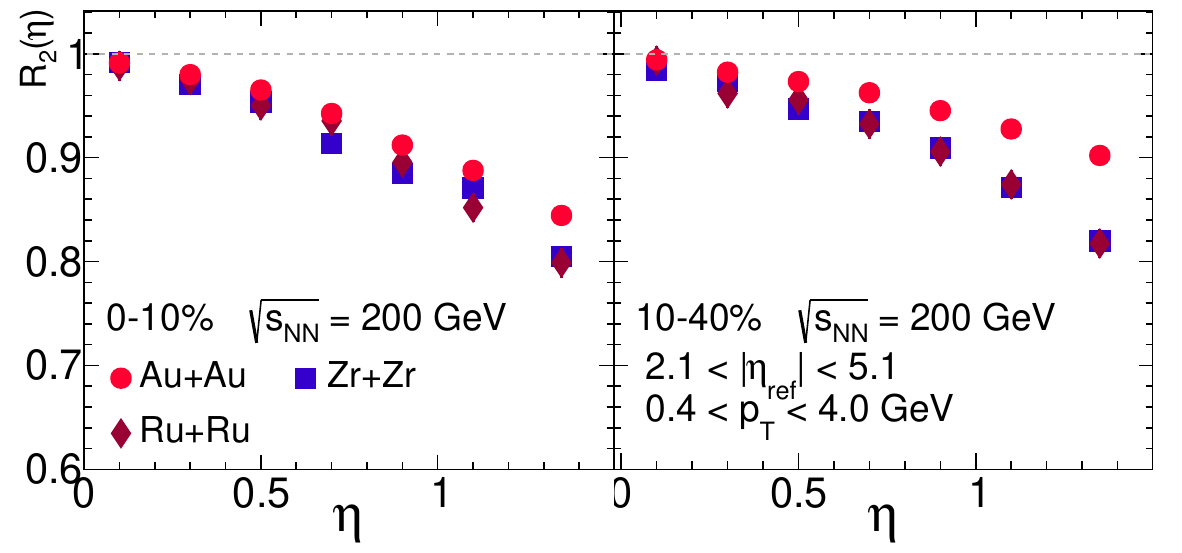}
    \caption{The $R_2(\eta)$ compared at \sqsnn = 200 GeV in Au+Au (orange), Zr+Zr (red) and Ru+Ru (blue) collisions in centrality 0-10\% and 10-40\%. The vertical line on the data points represent the statistical uncertainty}
    \label{fig_R2System}
\end{figure}

To gain insight into collision energy and system size dependence of event plane twist effect of longitudinal flow decorrelation. The $R_2(\eta)$ is firstly compared in Au+Au collisions at \sqsnn = 19.6, 27, 54.4, 200 GeV and isobar (Zr+Zr and Ru+Ru) collisions at \sqsnn = 200 GeV as shown in Figure \ref{fig_R2Energy} and Figure \ref{fig_R2System}. Lower energy has larger event plane twist effect in 10-40\% centrality. This significant collision energy dependence due to lower energy becomes less boost invariant. In addition, $R_2(\eta)$ shows a significant system size dependence which is similar as $r_2(\eta)$. The statistical precision required for a reliable $R_3(\eta)$ analysis in this analysis is prohibitively high and precludes its calculation with the current dataset.
%=======================================
% Figure7: R2 collision energy dependence
%=======================================
%=======================================
% Figure8: R2 system size dependence
%=======================================
%=======================================
%%%%%%%%%%%%%%%%%%%%%%%%%%%%%%%%%%%%%%%%%%%%%%%%%%%%%%%%%%%%%%%%%%%%%%%%%%%%%%%%%%%
%%%%%%%%%%%%%%%%%%%%%%%%%%%%%%%%%%%%%%%%%%%%%%%%%%%%%%%%%%%%%%%%%%%%%%%%%%%%%%%%%%%
\section{Summary}
We investigated the collision energy and the system size dependence of longitudinal flow decorrelation using the AMPT model for Au+Au collisions at \sqsnn = 19.6, 27, 54.4, and 200 GeV, as well as for isobar (Zr+Zr and Ru+Ru) collisions at \sqsnn = 200 GeV. The longitudinal flow decorrelation is characterized by $r_n(\eta)$, which captures the combined effects of $v_n$ asymmetry and event plane twist, whereas $R_n(\eta)$ solely isolates the contribution of the event plane twist. The slope parameter $F_{n}$ (n = 2, 3) as a function of $N_{\mathrm{part}}$ and the scaled system size $N_{\mathrm{part}}/2A$ to quantify the decorrelation strength. The results show that both $r_n(\eta)$ and $R_n(\eta)$ exhibit a clear collision energy and system size dependence. Specifically, lower energies have stronger longitudinal decorrelation even after beam rapidity scaling, which is consistent with the less boost invariant picture at lower collision energies. Notably, An opposite trend of system size dependence between $F_2$ and $F_3$ as a function of $N_{\mathrm{part}}$ is observed, which is consistent with the previous studies. However, the opposite trend disappears when the scaled system size $N_{\mathrm{part}}/2A$ is considered, which needs to be verified by RHIC-STAR experimental data. These systematic studies of the collision energy and system size dependence provide the most stringent constraints to date on the initial state geometry and subsequent dynamical evolution in relativistic heavy ion collisions.
%%%%%%%%%%%%%%%%%%%%%%%%%%%%%%%%%%%%%%%%%%%%%%%%%%%%%%%%%%%%%%%%%%%%%%%%%%%%%%%%%%%
%%%%%%%%%%%%%%%%%%%%%%%%%%%%%%%%%%%%%%%%%%%%%%%%%%%%%%%%%%%%%%%%%%%%%%%%%%%%%%%%%%%
\section{Acknowledgments}
We thank Jianing Dong for maintaining the high-quality performance of the computer facility. M. Nie, L.Yi and Z. Chen are supported by the National Natural Science Foundation of China under Grant No. 12105156, No. 11890710 and No. 11890713, National Key R\&D Program of China under Grant No. 2022YFA1604903 and Shandong Provincial Natural Science Foundation under Grant No. ZR2021QA084. J. Jia is supported by DOE Award No. DEFG0287ER40331.
\bibliography{references}

@article{Ollitrault:1992bk,
    author = "Ollitrault, Jean-Yves",
    title = "{Anisotropy as a signature of transverse collective flow}",
    reportNumber = "SACLAY-SPH-T-92-016",
    doi = "10.1103/PhysRevD.46.229",
    journal = "Phys. Rev. D",
    volume = "46",
    pages = "229--245",
    year = "1992"
}

@article{Ollitrault:1993ba,
    author = "Ollitrault, Jean-Yves",
    title = "{Determination of the reaction plane in ultrarelativistic nuclear collisions}",
    eprint = "hep-ph/9303247",
    archivePrefix = "arXiv",
    reportNumber = "SACLAY-SPH-T-93-026",
    doi = "10.1103/PhysRevD.48.1132",
    journal = "Phys. Rev. D",
    volume = "48",
    pages = "1132--1139",
    year = "1993"
}

@article{Voloshin:1994mz,
    author = "Voloshin, S. and Zhang, Y.",
    title = "{Flow study in relativistic nuclear collisions by Fourier expansion of Azimuthal particle distributions}",
    eprint = "hep-ph/9407282",
    archivePrefix = "arXiv",
    doi = "10.1007/s002880050141",
    journal = "Z. Phys. C",
    volume = "70",
    pages = "665--672",
    year = "1996"
}

@article{Poskanzer:1998yz,
    author = "Poskanzer, Arthur M. and Voloshin, S. A.",
    title = "{Methods for analyzing anisotropic flow in relativistic nuclear collisions}",
    eprint = "nucl-ex/9805001",
    archivePrefix = "arXiv",
    doi = "10.1103/PhysRevC.58.1671",
    journal = "Phys. Rev. C",
    volume = "58",
    pages = "1671--1678",
    year = "1998"
}

@article{STAR:2000ekf,
    author = "Ackermann, K. H. and others",
    collaboration = "STAR",
    title = "{Elliptic flow in Au + Au collisions at (S(NN))**(1/2) = 130 GeV}",
    eprint = "nucl-ex/0009011",
    archivePrefix = "arXiv",
    doi = "10.1103/PhysRevLett.86.402",
    journal = "Phys. Rev. Lett.",
    volume = "86",
    pages = "402--407",
    year = "2001"
}

@article{STAR:2004jwm,
    author = "Adams, J. and others",
    collaboration = "STAR",
    title = "{Azimuthal anisotropy in Au+Au collisions at s(NN)**(1/2) = 200-GeV}",
    eprint = "nucl-ex/0409033",
    archivePrefix = "arXiv",
    doi = "10.1103/PhysRevC.72.014904",
    journal = "Phys. Rev. C",
    volume = "72",
    pages = "014904",
    year = "2005"
}

@article{STAR:2013qio,
    author = "Adamczyk, L. and others",
    collaboration = "STAR",
    title = "{Third Harmonic Flow of Charged Particles in Au+Au Collisions at sqrtsNN = 200 GeV}",
    eprint = "1301.2187",
    archivePrefix = "arXiv",
    primaryClass = "nucl-ex",
    reportNumber = "LBNL (2013), LBNL-(2013)",
    doi = "10.1103/PhysRevC.88.014904",
    journal = "Phys. Rev. C",
    volume = "88",
    number = "1",
    pages = "014904",
    year = "2013"
}

@article{PHENIX:2004jsa,
    author = "Adler, S. S. and others",
    collaboration = "PHENIX",
    title = "{Saturation of azimuthal anisotropy in Au + Au collisions at s(NN)**(1/2) 62-GeV to 200-GeV}",
    eprint = "nucl-ex/0411040",
    archivePrefix = "arXiv",
    doi = "10.1103/PhysRevLett.94.232302",
    journal = "Phys. Rev. Lett.",
    volume = "94",
    pages = "232302",
    year = "2005"
}

@article{PHOBOS:2006dbo,
    author = "Alver, B. and others",
    collaboration = "PHOBOS",
    title = "{System size, energy, pseudorapidity, and centrality dependence of elliptic flow}",
    eprint = "nucl-ex/0610037",
    archivePrefix = "arXiv",
    doi = "10.1103/PhysRevLett.98.242302",
    journal = "Phys. Rev. Lett.",
    volume = "98",
    pages = "242302",
    year = "2007"
}

@article{ALICE:2010suc,
    author = "Aamodt, K and others",
    collaboration = "ALICE",
    title = "{Elliptic flow of charged particles in Pb-Pb collisions at 2.76 TeV}",
    eprint = "1011.3914",
    archivePrefix = "arXiv",
    primaryClass = "nucl-ex",
    reportNumber = "CERN-PH-EP-2010-059",
    doi = "10.1103/PhysRevLett.105.252302",
    journal = "Phys. Rev. Lett.",
    volume = "105",
    pages = "252302",
    year = "2010"
}

@article{ALICE:2011ab,
    author = "Aamodt, K. and others",
    collaboration = "ALICE",
    title = "{Higher harmonic anisotropic flow measurements of charged particles in Pb-Pb collisions at $\sqrt{s_{NN}}$=2.76 TeV}",
    eprint = "1105.3865",
    archivePrefix = "arXiv",
    primaryClass = "nucl-ex",
    reportNumber = "CERN-PH-EP-2011-073",
    doi = "10.1103/PhysRevLett.107.032301",
    journal = "Phys. Rev. Lett.",
    volume = "107",
    pages = "032301",
    year = "2011"
}

@article{ATLAS:2012at,
    author = "Aad, Georges and others",
    collaboration = "ATLAS",
    title = "{Measurement of the azimuthal anisotropy for charged particle production in $\sqrt{s_{NN}}=2.76$ TeV lead-lead collisions with the ATLAS detector}",
    eprint = "1203.3087",
    archivePrefix = "arXiv",
    primaryClass = "hep-ex",
    reportNumber = "CERN-PH-EP-2012-035",
    doi = "10.1103/PhysRevC.86.014907",
    journal = "Phys. Rev. C",
    volume = "86",
    pages = "014907",
    year = "2012"
}

@article{CMS:2013wjq,
    author = "Chatrchyan, Serguei and others",
    collaboration = "CMS",
    title = "{Measurement of Higher-Order Harmonic Azimuthal Anisotropy in PbPb Collisions at $\sqrt{s_{NN}}$ = 2.76 TeV}",
    eprint = "1310.8651",
    archivePrefix = "arXiv",
    primaryClass = "nucl-ex",
    reportNumber = "CMS-HIN-11-005, CERN-PH-EP-2013-196",
    doi = "10.1103/PhysRevC.89.044906",
    journal = "Phys. Rev. C",
    volume = "89",
    number = "4",
    pages = "044906",
    year = "2014"
}

@article{Nie:2019bgd,
    author = "Nie, Maowu",
    collaboration = "STAR",
    title = "{Measurement of longitudinal decorrelation of anisotropic flow $v_2$ and $v_3$ in 200 GeV Au+Au collisions at STAR}",
    doi = "10.1016/j.nuclphysa.2018.09.068",
    journal = "Nucl. Phys. A",
    volume = "982",
    pages = "403--406",
    year = "2019"
}

@article{Yan:2023ugh,
    author = "Yan, Gaoguo",
    collaboration = "STAR",
    title = "{Probing Initial- and Final-state Effects of Heavy-ion Collisions with STAR Experiment}",
    doi = "10.5506/APhysPolBSupp.16.1-A137",
    journal = "Acta Phys. Polon. Supp.",
    volume = "16",
    number = "1",
    pages = "1--A137",
    year = "2023"
}

@article{CMS:2015xmx,
    author = "Khachatryan, Vardan and others",
    collaboration = "CMS",
    title = "{Evidence for transverse momentum and pseudorapidity dependent event plane fluctuations in PbPb and pPb collisions}",
    eprint = "1503.01692",
    archivePrefix = "arXiv",
    primaryClass = "nucl-ex",
    reportNumber = "CMS-HIN-14-012, CERN-PH-EP-2015-039",
    doi = "10.1103/PhysRevC.92.034911",
    journal = "Phys. Rev. C",
    volume = "92",
    number = "3",
    pages = "034911",
    year = "2015"
}

@article{ATLAS:2017rij,
    author = "Aaboud, Morad and others",
    collaboration = "ATLAS",
    title = "{Measurement of longitudinal flow decorrelations in Pb+Pb collisions at $\sqrt{s_{\text {NN}}}=2.76$ and 5.02 TeV with the ATLAS detector}",
    eprint = "1709.02301",
    archivePrefix = "arXiv",
    primaryClass = "nucl-ex",
    reportNumber = "CERN-EP-2017-191",
    doi = "10.1140/epjc/s10052-018-5605-7",
    journal = "Eur. Phys. J. C",
    volume = "78",
    number = "2",
    pages = "142",
    year = "2018"
}

@article{ATLAS:2020sgl,
    author = "Aad, Georges and others",
    collaboration = "ATLAS",
    title = "{Longitudinal Flow Decorrelations in Xe+Xe Collisions at $\sqrt{s_{\mathrm{NN}}}=5.44$  TeV with the ATLAS Detector}",
    eprint = "2001.04201",
    archivePrefix = "arXiv",
    primaryClass = "nucl-ex",
    reportNumber = "CERN-EP-2019-275",
    doi = "10.1103/PhysRevLett.126.122301",
    journal = "Phys. Rev. Lett.",
    volume = "126",
    number = "12",
    pages = "122301",
    year = "2021"
}

@article{Bozek:2010vz,
    author = "Bozek, Piotr and Broniowski, Wojciech and Moreira, Joao",
    title = "{Torqued fireballs in relativistic heavy-ion collisions}",
    eprint = "1011.3354",
    archivePrefix = "arXiv",
    primaryClass = "nucl-th",
    doi = "10.1103/PhysRevC.83.034911",
    journal = "Phys. Rev. C",
    volume = "83",
    pages = "034911",
    year = "2011"
}

@article{Xiao:2012uw,
    author = "Xiao, Kai and Liu, Feng and Wang, Fuqiang",
    title = "{Event-plane decorrelation over pseudorapidity and its effect on azimuthal anisotropy measurements in relativistic heavy-ion collisions}",
    eprint = "1208.1195",
    archivePrefix = "arXiv",
    primaryClass = "nucl-th",
    doi = "10.1103/PhysRevC.87.011901",
    journal = "Phys. Rev. C",
    volume = "87",
    number = "1",
    pages = "011901",
    year = "2013"
}

@article{Jia:2014ysa,
    author = "Jia, Jiangyong and Huo, Peng",
    title = "{Forward-backward eccentricity and participant-plane angle fluctuations and their influences on longitudinal dynamics of collective flow}",
    eprint = "1403.6077",
    archivePrefix = "arXiv",
    primaryClass = "nucl-th",
    doi = "10.1103/PhysRevC.90.034915",
    journal = "Phys. Rev. C",
    volume = "90",
    number = "3",
    pages = "034915",
    year = "2014"
}

@article{Bozek:2015bna,
    author = "Bozek, Piotr and Broniowski, Wojciech",
    title = "{The torque effect and fluctuations of entropy deposition in rapidity in ultra-relativistic nuclear collisions}",
    eprint = "1506.02817",
    archivePrefix = "arXiv",
    primaryClass = "nucl-th",
    doi = "10.1016/j.physletb.2015.11.054",
    journal = "Phys. Lett. B",
    volume = "752",
    pages = "206--211",
    year = "2016"
}

@article{Pang:2015zrq,
    author = "Pang, Long-Gang and Petersen, Hannah and Qin, Guang-You and Roy, Victor and Wang, Xin-Nian",
    title = "{Decorrelation of anisotropic flow along the longitudinal direction}",
    eprint = "1511.04131",
    archivePrefix = "arXiv",
    primaryClass = "nucl-th",
    doi = "10.1140/epja/i2016-16097-x",
    journal = "Eur. Phys. J. A",
    volume = "52",
    number = "4",
    pages = "97",
    year = "2016"
}

@article{Schenke:2016ksl,
    author = "Schenke, Bjoern and Schlichting, Soeren",
    title = "{3D glasma initial state for relativistic heavy ion collisions}",
    eprint = "1605.07158",
    archivePrefix = "arXiv",
    primaryClass = "hep-ph",
    doi = "10.1103/PhysRevC.94.044907",
    journal = "Phys. Rev. C",
    volume = "94",
    number = "4",
    pages = "044907",
    year = "2016"
}

@article{Shen:2017bsr,
    author = {Shen, Chun and Schenke, Bj\"orn},
    title = "{Dynamical initial state model for relativistic heavy-ion collisions}",
    eprint = "1710.00881",
    archivePrefix = "arXiv",
    primaryClass = "nucl-th",
    doi = "10.1103/PhysRevC.97.024907",
    journal = "Phys. Rev. C",
    volume = "97",
    number = "2",
    pages = "024907",
    year = "2018"
}

@article{Wu:2018cpc,
    author = "Wu, Xiang-Yu and Pang, Long-Gang and Qin, Guang-You and Wang, Xin-Nian",
    title = "{Longitudinal fluctuations and decorrelations of anisotropic flows at energies available at the CERN Large Hadron Collider and at the BNL Relativistic Heavy Ion Collider}",
    eprint = "1805.03762",
    archivePrefix = "arXiv",
    primaryClass = "nucl-th",
    doi = "10.1103/PhysRevC.98.024913",
    journal = "Phys. Rev. C",
    volume = "98",
    number = "2",
    pages = "024913",
    year = "2018"
}

@article{Jia:2017kdq,
    author = "Jia, Jiangyong and Huo, Peng and Ma, Guoliang and Nie, Maowu",
    title = "{Observables for longitudinal flow correlations in heavy-ion collisions}",
    eprint = "1701.02183",
    archivePrefix = "arXiv",
    primaryClass = "nucl-th",
    doi = "10.1088/1361-6471/aa74c3",
    journal = "J. Phys. G",
    volume = "44",
    number = "7",
    pages = "075106",
    year = "2017"
}

@article{Miller:2007ri,
  title = {Glauber Modeling in High Energy Nuclear Collisions},
  author = {Miller, Michael L. and Reygers, Klaus and Sanders, Stephen J. and Steinberg, Peter},
  journal = {Annu. Rev. Nucl. Part. Sci.},
  volume = {57},
  pages = {205-243},
  year = {2007},
  doi = {10.1146/annurev.nucl.57.090506.123020},
  archiveprefix = {arXiv}
}

@article{Lin:2004en,
    author = "Lin, Zi-Wei and Ko, Che Ming and Li, Bao-An and Zhang, Bin and Pal, Subrata",
    title = "{A Multi-phase transport model for relativistic heavy ion collisions}",
    eprint = "nucl-th/0411110",
    archivePrefix = "arXiv",
    doi = "10.1103/PhysRevC.72.064901",
    journal = "Phys. Rev. C",
    volume = "72",
    pages = "064901",
    year = "2005"
}

@article{Wang:1991hta,
    author = "Wang, Xin-Nian and Gyulassy, Miklos",
    title = "{HIJING: A Monte Carlo model for multiple jet production in p p, p A and A A collisions}",
    reportNumber = "LBL-31036",
    doi = "10.1103/PhysRevD.44.3501",
    journal = "Phys. Rev. D",
    volume = "44",
    pages = "3501--3516",
    year = "1991"
}

@article{Zhang:1997ej,
    author = "Zhang, Bin",
    title = "{ZPC 1.0.1: A Parton cascade for ultrarelativistic heavy ion collisions}",
    eprint = "nucl-th/9709009",
    archivePrefix = "arXiv",
    reportNumber = "CU-TP-853",
    doi = "10.1016/S0010-4655(98)00010-1",
    journal = "Comput. Phys. Commun.",
    volume = "109",
    pages = "193--206",
    year = "1998"
}

@article{Li:1995pra,
    author = "Li, Bao-An and Ko, Che Ming",
    title = "{Formation of superdense hadronic matter in high-energy heavy ion collisions}",
    eprint = "nucl-th/9505016",
    archivePrefix = "arXiv",
    doi = "10.1103/PhysRevC.52.2037",
    journal = "Phys. Rev. C",
    volume = "52",
    pages = "2037--2063",
    year = "1995"
}

@article{Ma:2016fve,
    author = "Ma, Guo-Liang and Lin, Zi-Wei",
    title = "{Predictions for $\sqrt {s_{NN}}=5.02$ TeV Pb+Pb Collisions from a Multi-Phase Transport Model}",
    eprint = "1601.08160",
    archivePrefix = "arXiv",
    primaryClass = "nucl-th",
    doi = "10.1103/PhysRevC.93.054911",
    journal = "Phys. Rev. C",
    volume = "93",
    number = "5",
    pages = "054911",
    year = "2016"
}

@article{Dixit:2023bho,
    author = "Dixit, Prabhupada and Nasim, Md.",
    title = "{Longitudinal flow decorrelation in heavy-ion collision at RHIC energies using a multi-phase transport model}",
    eprint = "2307.08406",
    archivePrefix = "arXiv",
    primaryClass = "hep-ph",
    journal = " ",
    month = "7",
    year = "2023"
}

@article{Jia:2024xvl,
    author = "Jia, Jiangyong and Huang, Shengli and Zhang, Chunjian and Bhatta, Somadutta",
    title = "{Sources of longitudinal flow decorrelations in high-energy nuclear collisions}",
    eprint = "2408.15006",
    archivePrefix = "arXiv",
    primaryClass = "nucl-th",
    journal = " ",
    month = "8",
    year = "2024"
}

@article{Nie:2022gbg,
    author = "Nie, Maowu and Zhang, Chunjian and Chen, Zhenyu and Yi, Li and Jia, Jiangyong",
    title = "{Impact of nuclear structure on longitudinal flow decorrelations in high-energy isobar collisions}",
    eprint = "2208.05416",
    archivePrefix = "arXiv",
    primaryClass = "nucl-th",
    doi = "10.1016/j.physletb.2023.138177",
    journal = "Phys. Lett. B",
    volume = "845",
    pages = "138177",
    year = "2023"
}

@article{PHOBOS:2004nvy,
    author = "Back, B. B. and others",
    collaboration = "PHOBOS",
    title = "{Energy dependence of elliptic flow over a large pseudorapidity range in Au+Au collisions at RHIC}",
    eprint = "nucl-ex/0406021",
    archivePrefix = "arXiv",
    doi = "10.1103/PhysRevLett.94.122303",
    journal = "Phys. Rev. Lett.",
    volume = "94",
    pages = "122303",
    year = "2005"
}

@article{Gale:2013da,
	author = "Gale, Charles and Jeon, Sangyong and Schenke, Bjoern",
	title = "{Hydrodynamic Modeling of Heavy-Ion Collisions}",
	eprint = "1301.5893",
	archivePrefix = "arXiv",
	primaryClass = "nucl-th",
	doi = "10.1142/S0217751X13400113",
	journal = "Int. J. Mod. Phys. A",
	volume = "28",
	pages = "1340011",
	year = "2013"
}

@article{Denicol:2015nhu,
	author = "Denicol, Gabriel and Monnai, Akihiko and Schenke, Bjoern",
	title = "{Moving forward to constrain the shear viscosity of QCD matter}",
	eprint = "1512.01538",
	archivePrefix = "arXiv",
	primaryClass = "nucl-th",
	doi = "10.1103/PhysRevLett.116.212301",
	journal = "Phys. Rev. Lett.",
	volume = "116",
	number = "21",
	pages = "212301",
	year = "2016"
}

@article{Huo:2013qma,
	author = "Huo, Peng and Jia, Jiangyong and Mohapatra, Soumya",
	title = "{Elucidating the event-by-event flow fluctuations in heavy-ion collisions via the event shape selection technique}",
	eprint = "1311.7091",
	archivePrefix = "arXiv",
	primaryClass = "nucl-ex",
	doi = "10.1103/PhysRevC.90.024910",
	journal = "Phys. Rev. C",
	volume = "90",
	number = "2",
	pages = "024910",
	year = "2014"
}

@article{Bozek:2015bha,
	author = "Bo\.zek, Piotr and Broniowski, Wojciech and Olszewski, Adam",
	title = "{Hydrodynamic modeling of pseudorapidity flow correlations in relativistic heavy-ion collisions and the torque effect}",
	eprint = "1503.07425",
	archivePrefix = "arXiv",
	primaryClass = "nucl-th",
	doi = "10.1103/PhysRevC.91.054912",
	journal = "Phys. Rev. C",
	volume = "91",
	pages = "054912",
	year = "2015"
}

@article{Behera:2020mol,
    author = "Behera, Arabinda and Nie, Maowu and Jia, Jiangyong",
    title = "{Longitudinal eccentricity decorrelations in heavy ion collisions}",
    eprint = "2003.04340",
    archivePrefix = "arXiv",
    primaryClass = "nucl-th",
    doi = "10.1103/PhysRevResearch.2.023362",
    journal = "Phys. Rev. Res.",
    volume = "2",
    number = "2",
    pages = "023362",
    year = "2020"
}

\end{document}